\documentclass[10pt,twocolumn]{article}
\usepackage{sp_template}

\usepackage{bm}
\usepackage{nicefrac}
\usepackage{adjustbox}
\usepackage{tikz}
\usetikzlibrary{matrix,positioning,fit}

\definecolor{coralPink}{HTML}{FF858D}

\newcommand{\sierpinski}[2][]{\tikz[#1]{
  \draw[fill=black] rectangle(1,1);
  \foreach \n[
      evaluate=\n as \m using \n-1,
      evaluate=\n as \s using 1/3^\n,
      evaluate=\m as \p using 3^\m] in {1,...,#2}{
    \foreach \k[evaluate=\k as \x using (2*\k-1)/2/3^\m] in {1,...,\p}{
      \foreach \j[evaluate=\j as \y using (2*\j-1)/2/3^\m] in {1,...,\p}{
        \node[fill=white, minimum size=\s cm, inner sep=0] at (\x,\y){};
}}}}}

\DeclareMathOperator\arctanh{arctanh}
\newcommand{\rme}{{\rm e}}
\newcommand{\rmi}{{\rm i}}

\hypersetup{
  pdftitle  = {Critical Temperature(s) \\ of Sierpinski Carpet(s)},
  pdfauthor = {R. Ben Ali Zinati, G. Gori, A. Codello},
}

\title{Critical Temperature(s) \\ of Sierpi\'{n}ski Carpet(s)}

\author{%
  R. Ben Al\`i Zinati\affilmark{1}%
  \and
  G. Gori\affilmark{2,3}%
  \and
  A. Codello\affilmark{4,5}%
}

\affiliation[1]{Universit\`a della Svizzera italiana, 6900 Lugano, Switzerland}
\affiliation[2]{CNR-INO, Area Science Park, Basovizza, 34149 Trieste, Italy}
\affiliation[3]{Institut f\"ur Theoretische Physik, Universit\"at Heidelberg, 69120 Heidelberg, Germany}
\affiliation[4]{DSMN, Ca'\ Foscari University of Venice, Via Torino 155, 30172 Venice, Italy}
\affiliation[5]{IFFI, Universidad de la Rep\'ublica, J.H.y Reissig 565, 11300 Montevideo, Uruguay}

\date{\today}
\preprint{https://doi.org/10.1140/epjb/s10051-026-01196-1}
\begin{document}

\twocolumn[{%
  \maketitle
  \begin{abstract}
We present a key algorithmic improvement to the generalized combinatorial Feynman--Vdovichenko method for calculating the critical temperature of the Ising model on Sierpi\'{n}ski carpets $SC_k(a,b)$, originally introduced in {\tt arxiv:1505.02699}.
By reformulating the method in terms of purely real-valued transition matrices, we substantially reduce their dimension.
This optimization, together with modern computational resources, enables us to reach generation $k=10$ for the canonical $SC_k(3,1)$ carpet.
Extrapolation from these data yields the most accurate estimate to date of the critical temperature $T_c^{(3,1)} =  1.4782927(26)$.
We further extend the analysis to additional members of the $SC_k(a,b)$ family and report their corresponding critical temperatures.
  \end{abstract}
}]

%───────────────────────────────────────────────────────────────────────────────
\section*{Introduction}
%───────────────────────────────────────────────────────────────────────────────

The Sierpi\'{n}ski carpet family $SC_k(a,b)$ represents a central class of fractals with infinite ramification number \cite{gefen:1980aa,gefen_1984aa}, where the Ising model is characterized by a non-zero critical temperature $T_c$ \cite{vezzani:2003aa, Shinoda_2002}.
However, the precise determination of $T_c$ is notoriously challenging and conventional Monte Carlo simulations suffer from severe slow convergence \cite{Bonnier:1987aa, Monceau:1998aa, Carmona:1998aa, Monceau:2001aa, Pruessner:2001aa, Monceau_2004, Bab:2005aa,BAB2009370,Jang_2024}.
The other classical approach is real space RG \cite{Bonnier_1988, Hsiao}, which in recent years have been significantly refined through the development of (high-order) tensor renormalization group techniques \cite{Genzor2016, Genzor2019, Genzor2023}.
A major advance was achieved with the introduction of a generalized combinatorial Feynman-Vdovichenko method \cite{Codello:2015bia}, which remains the most efficient approach available to date. This technique maps the determination of the critical temperature onto finding the real eigenvalue $\lambda_c>1$ of a transition matrix $\mathbb{W}_k $, obtained numerically via the Arnoldi algorithm. The method's efficacy, later independently rediscovered \cite{Perreau}, is nonetheless hampered by its computational scaling. The matrix $\mathbb{W}_k$ grows rapidly with the fractal generation $k$; for the $SC_k(3,1)$ carpet, its size increases eightfold at each iteration, with an even more severe growth for other members of the family.

\begin{figure}[ht!]
\centering
\begin{adjustbox}{width=\columnwidth}
\begin{tikzpicture}[font=\footnotesize]
\pgfdeclarelayer{background}
\pgfsetlayers{background,main}

%\node (title) at (0,0) {\textbf{}};

\matrix (M) [matrix of nodes,
            nodes={draw, minimum width=0.7cm, minimum height=0.45cm},
            row sep=0pt, column sep=0pt]
            %below=0pt of title]
{
|[name=A, fill=coralPink!25]| $(3,1)$
  & $(4,2)$ & $(5,3)$ & $(5,1)$ & $(6,4)$ & $(6,2)$ & $(7,5)$ & $(7,3)$ & $(7,1)$ \\
};

\node (ls) [below=10pt of M] {};

\matrix (C) [
   matrix of nodes,
   column sep=10pt,
   row sep=0pt,
   below=20pt of ls
]
{
   |[name=C1]| \sierpinski{1}  &
   |[name=C2]| \sierpinski{2}  &
   |[name=C3]| \sierpinski{3}  &
   |[name=C4]| \sierpinski{4}  &
   |[name=C5]| {\raisebox{3.1ex}{~\Large $\cdots$}} \\
};

\node (L1) [above=2pt of C1] {$SC_1(3,1)$};
\node (L2) [above=2pt of C2] {$SC_2(3,1)$};
\node (L3) [above=2pt of C3] {$SC_3(3,1)$};
\node (L4) [above=2pt of C4] {$SC_4(3,1)$};
\node (L5) [above=17pt of C5] {$~\dots$};

\node[
   draw,
   coralPink,
   inner sep=5pt,
   name=Sbox,
   rounded corners,
   fit=(L1)(L5)(C1)(C5)
] {};

\draw[rounded corners, coralPink, -stealth]
  (A.south) -- ++(0,-8pt) -| (Sbox.north);

\begin{pgfonlayer}{background}
  \node[
    draw,
    rounded corners,
    fill=coralPink!25,
    inner sep=1pt,
    fit=(C3),
    name=C3box
  ] {};
\end{pgfonlayer}

\node (img) [below=15pt of C3]
  {\includegraphics[width=\columnwidth]{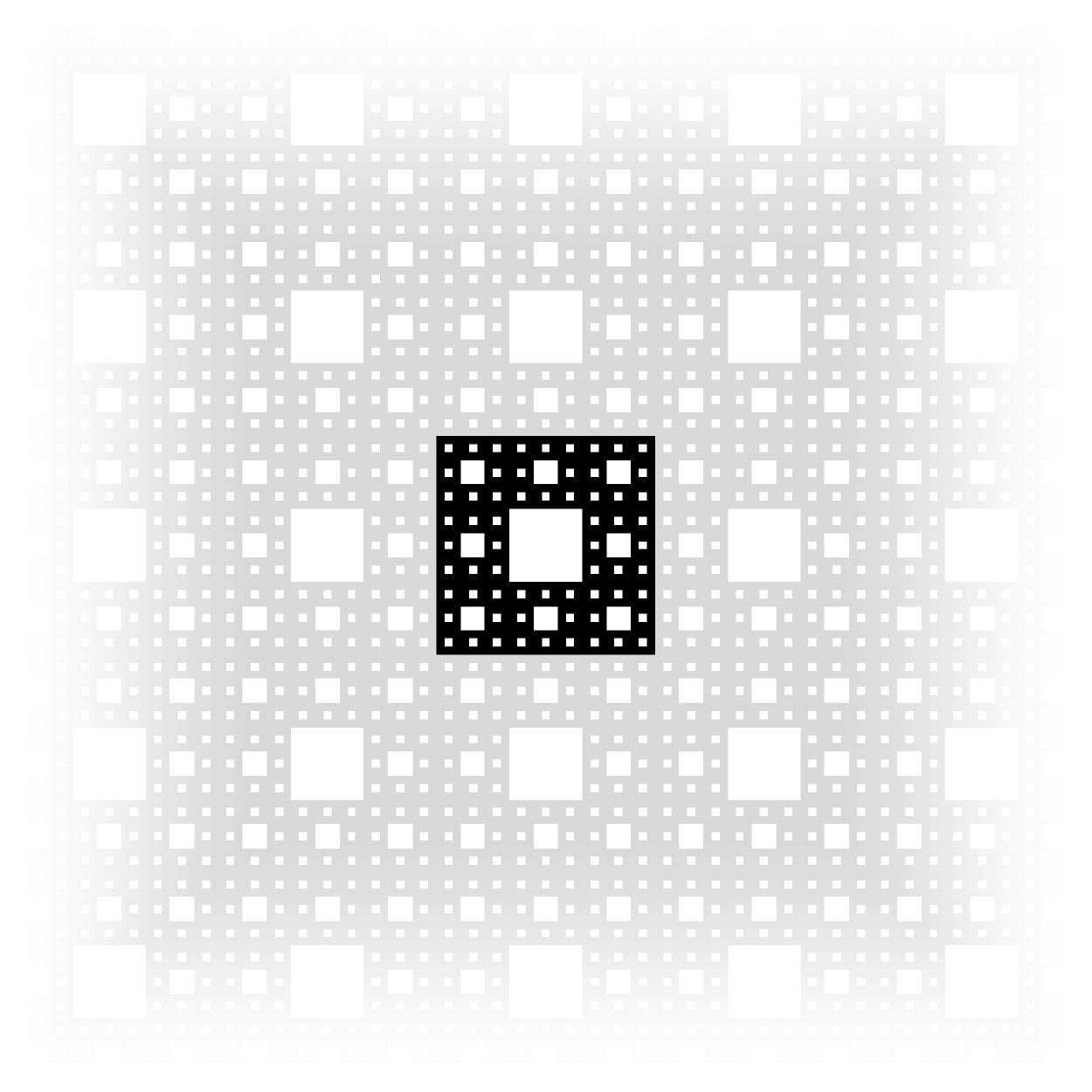}};
\draw[rounded corners, coralPink, -stealth]
 (C3box.south) -- ++(0,-90pt);

\end{tikzpicture}
\end{adjustbox}
\caption{Construction pipeline for Sierpi\'{n}ski carpets $SC_k(a,b)$.
As an example, the canonical case $(3,1)$ is highlighted together with its first finite iterations (middle row).
The bottom panel illustrates the tiling of the plane using the selected approximation $SC_3(3,1)$. Convergence to the full carpet is obtained in the limit $k\to \infty$.
}
\label{carpet}
\end{figure}
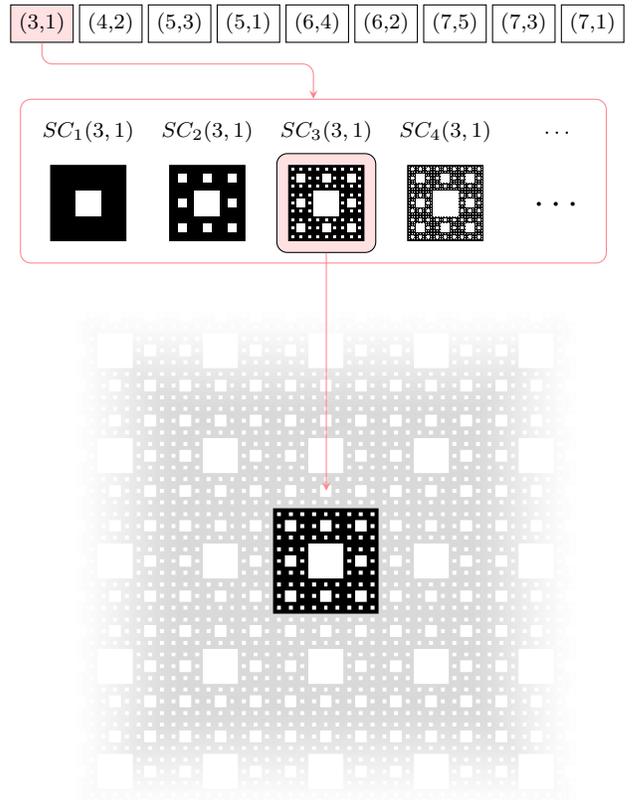

In this work, we build on the method introduced in Ref.~\cite{Codello:2015bia} by reformulating the construction in a way that renders the associated transition matrices purely real, with entries restricted to $\pm 1$.
While leaving the spectrum unchanged, this modification reduces the matrix dimension by a factor of two.
Combined with the increase in available computational resources over the past decade, this refinement enables us to reach generation $k=10$ for the canonical Sierpi\'{n}ski carpet $SC_k(3,1)$, well beyond previously accessible limits.
Extrapolation from data obtained up to this generation yields the most accurate estimates of critical temperature to date, and we further extend the analysis to other carpets within the $SC_k(a,b)$ family.
In addition to the canonical $(3,1)$, we investigate the carpets $(4,2)$, $(5,3)$, $(5,1)$, $(6,4)$, $(6,2)$, and $(7,5)$, $(7,3)$, $(7,1)$.
The goal is to collect data across different values of $(a,b)$ and to explore possible patterns in the resulting critical behavior.

%───────────────────────────────────────────────────────────────────────────────
\section*{Improved Algorithm}
%───────────────────────────────────────────────────────────────────────────────

The method we employ is a generalization of the approach originally introduced by Vdovichenko \cite{Vodvicenko_1965}.
It reformulates the computation of the partition function as a combinatorial problem involving the enumeration of simply connected, closed contours on an infinite lattice.
This is achieved by constructing a transition matrix associated with the elementary plaquettes of the lattice.
The key ingredients of the method are: making the contours oriented and adding a nontrivial phase factor at each turn.
In the original formulation, a factor $\rme^{\rmi \pi / 4}$ ($\rme^{-\rmi \pi / 4}$) is assigned to each counterclockwise (clockwise) turn.
This choice ensures that each closed contour acquires an overall factor of $(-1)$, thereby canceling unwanted diagrams, as discussed in Ref.~\cite{Feynman_1972}.
While this choice of the phase factor is particularly symmetric it is definitely not unique: any assignment that produces a net factor of $(-1)$ per closed contour is equally valid.
In this work, we adopt the following modification: all turns carry a factor of $(+1)$, except for turns from the upward direction to the right and vice versa, which acquire a factor of $(-1)$.
The resulting transition matrix has the same spectrum as the original Feynman–Vdovichenko construction but is purely real, with entries restricted to $(+1)$, $(0)$, and $(-1)$.
This reformulation leads to a significant reduction in storage without affecting the numerical stability of the computation.
In particular, we verified through extensive numerical tests that the convergence properties of the implicitly restarted Arnoldi algorithm, as implemented in the ARPACK library \cite{ARPACK}, remain unchanged.
The resulting halving of the effective storage size is a crucial advantage for the present study, which aimed to access the largest possible fractal generations.

%───────────────────────────────────────────────────────────────────────────────
\section*{Results}
%───────────────────────────────────────────────────────────────────────────────

% SC31 — long table, full width alone
\begin{table}[t]
\centering\small
\caption{Data for the carpet $SC_k(3,1)$.}
\label{tableSC31}
\resizebox{\linewidth}{!}{%
\begin{tabular}{rccrrl}
\toprule
$k$ & $\lambda_c$ & $T_c$ & $L$ & \hphantom{5\,159\,780\,3} \# elem.\\
\midrule
2  & 1.8506569 & {\bf 1.6538550} & 9        & 636 \\
3  & 1.7746728 & {\bf 1.5675856} & 27       & 4\,980 \\
4  & 1.7381282 & {\bf 1.5256550} & 81       & 39\,516 \\
5  & 1.7197634 & {\bf 1.5044635} & 243      & 315\,156 \\
6  & 1.7101236 & {\bf 1.4933058} & 729      & 2\,518\,332 \\
7  & 1.7048483 & {\bf 1.4871895} & 2\,187   & 20\,137\,908 \\
8  & 1.7018490 & {\bf 1.4837088} & 6\,561   & 161\,077\,020 \\
9  & 1.7000852 & {\bf 1.4816609} & 19\,683  & 1\,288\,537\,428 \\
10 & 1.6990177 & {\bf 1.4804210} & 59\,049  & 10\,308\,063\,228 \\
\midrule
$\infty$ & 1.6971860(23) & {\bf 1.4782927(26)} & -- & -- \\
\bottomrule
\end{tabular}}
\end{table}

\begin{figure}[t]
\centering
\includegraphics[width=\columnwidth]{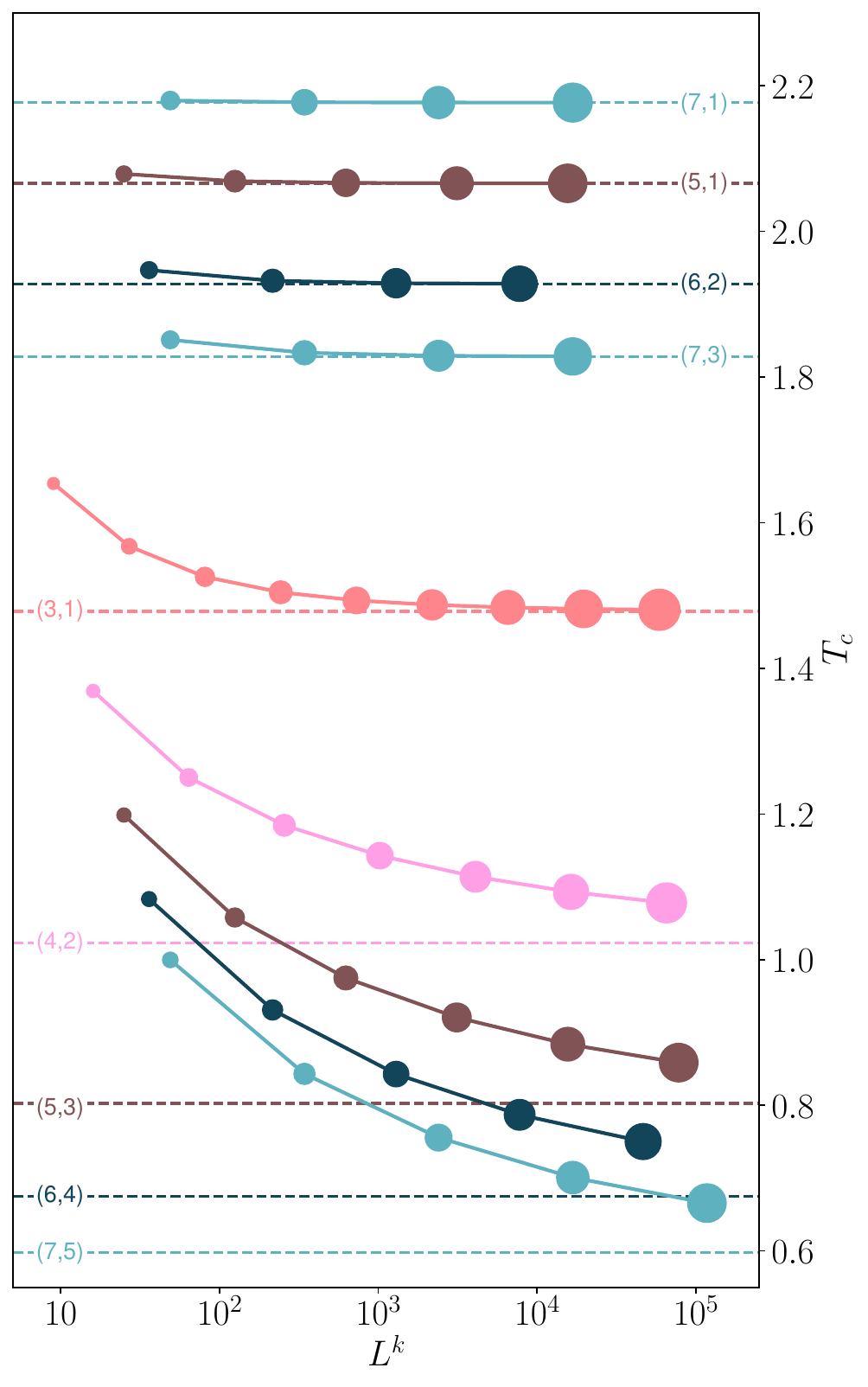}
\caption{Critical temperatures of the Sierpi\'{n}ski carpets $SC_k(a,b)$. $b$ decreases with $T_c$ while $k$ starts from two and goes up to the maximum value computed. Finally the size of the points is proportional to logarithm of the number of elements, highlighting the rapid growth in computational complexity.}
\label{comparison}
\end{figure}

The main critical quantities obtained from our technique are  the real critical eigenvalue $\lambda_c$
and the critical temperature $T_c=1/\arctanh(1/\lambda_c)$.
These quantities are computed as functions of the fractal generation $k$, together with the corresponding linear size $L$ and the number of nonzero entries of the transition matrix, for the various carpets $SC_k(a,b)$ considered.
The numerical results are reported in Tables~\ref{tableSC31}–\ref{tableSC75}.
For each carpet, at least four generations were computed, with the deepest coverage achieved for $SC_k(3,1)$, where we reach generation $k=10$.
At this size, the transition matrix contains approximately $10^{10}$ nonzero elements, representing an improvement of several orders of magnitude over previous combinatorial implementations.

The critical values obtained from our approach are essentially free from statistical uncertainty.
Residual errors arise only from floating-point arithmetic and from the intrinsic limitations of iterative eigensolvers.
These effects were carefully controlled by an appropriate choice of the shift parameter in the shift/invert Arnoldi method, ensuring robust convergence in all cases.
To extract the $k \to \infty$ critical values, we perform extrapolations based on the eigenvalue $\lambda_c$:
we fit the data for generations $k \in [k_{\min},k_{\max}]$, progressively increasing $k_{\min}$ up to $k_{\max}-4$, where $k_{\max}$ denotes the largest accessible generation.
The fitting function is taken to be exponential in the generation,
\begin{equation}
\lambda_c(k)=\lambda_c^\infty[k_{\min}]+\alpha\,\rme^{-\beta k}\,,
\end{equation}
which corresponds to a power-law correction in the system size.
The resulting estimates $\lambda_c^\infty[k_{\min}]$, together with their inferred errors, are then used to give our final estimate for $\lambda_c^\infty$ via a second exponential fit in a bootstrap-like fashion.
The preferred estimates are selected as those with the smallest uncertainties while retaining statistically excellent fit quality, with $p$-values always larger than $0.8$ and typically very close to unity.
For selected cases ---namely $SC_k(6,2)$, $SC_k(7,3)$, and $SC_k(7,1)$--- data are available only up to the $5^{th}$ generation.
In these cases, the final value of the extrapolated critical eigenvalue has been obtained fixing the exponent coefficient $\beta$ in the second-stage fit to an extrapolated value;
this procedure yields good fitting  but may slightly underestimate the final uncertainty.
Such cases are marked by an asterisk in Tables~\ref{tableSC62}, \ref{tableSC71}, and \ref{tableSC73}, where the quoted errors reach values as small as $4\times 10^{-8}$, potentially below the scale of typical numerical roundoff effects.
The choice to extrapolate the critical eigenvalue rather than the temperature itself is motivated by numerical stability: extrapolations performed directly in $T_c$ introduce nonanalytic behavior at low temperatures.
By contrast, the eigenvalue-based procedure performs optimally in known benchmark cases, including nearly one-dimensional systems (such as the Sierpi\'{n}ski gasket \cite{Codello:2015bia}) and nearly two-dimensional geometries, where it correctly captures both the presence or absence of long-range order and the location of the critical point.

For the fractal $SC_k(3,1)$ we provide for the first time data for the generations $k=9$ and $k=10$ (Table \ref{tableSC31}), extending the previous $k=7$ \cite{Codello:2015bia} and $k=8$ \cite{Perreau}. Our extrapolation gives $T_c^{(3,1)} = {\bf 1.4782927(26)}$,
which constitutes the most accurate estimate to date.
We note that this value is in excellent agreement with the higher-order tensor renormalization group estimate ($T_c^{(3,1)}$ = 1.47829) reported in Ref.~\cite{Genzor2019}.
%
%SC(5,1)
Among the other carpets studied, $SC_k(5,1)$ exhibits the fastest convergence.
As shown in Table~\ref{tableSC51}, the inclusion of generations $k=5$ and $k=6$ beyond \cite{Codello:2015bia} already fixes the critical temperature to three significant digits prior to extrapolation.
The final estimate, $T_c^{(5,1)} = {\bf 2.06602096(25)}$ confirms the general trend that carpets with fractal dimension closer to two---i.e., geometries more closely resembling the two-dimensional Ising model---display improved convergence properties.
The other three fractals that converge rapidly---the upper group of Fig. 2---are  $SC_k(6,2)$, $SC_k(7,1)$ and $SC_k(7,3)$. Their final temperatures---which can be read off from Tables~\ref{tableSC62}, \ref{tableSC71} and \ref{tableSC73}---are $T_c^{(6,2)} = {\bf  1.9279320(21)}$, $T_c^{(7,1)} = {\bf 2.1769455(6)}$ and $T_c^{(7,3)} = {\bf 1.82800090(4)}$.

For a carpet of generation $k$, the linear size of the fundamental tile is $L = a^k$.
The fractal dimension of this class of Sierpi\'{n}ski carpets follows from elementary considerations and is given by $$d_f = \frac{\log (a^2-b^2)}{\log a} \,.$$
In contrast, the spectral dimension $d_s$ is considerably more difficult to determine;
while one trivially has  $d_s \to 2$ in the limit $a>> b$, only general bounds are known in most cases. In particular, the general inequalities $d>d_f\geq d_s$ and $d_s > \tfrac{2d_f}{1+d_f}>1$ hold true \cite{Barlow}.
In this paper we thus study only the dependence $T_c(d_f)$ which is straightforward to determine from our data.

\begin{figure}[t]
\centering
\includegraphics[width=0.48\textwidth]{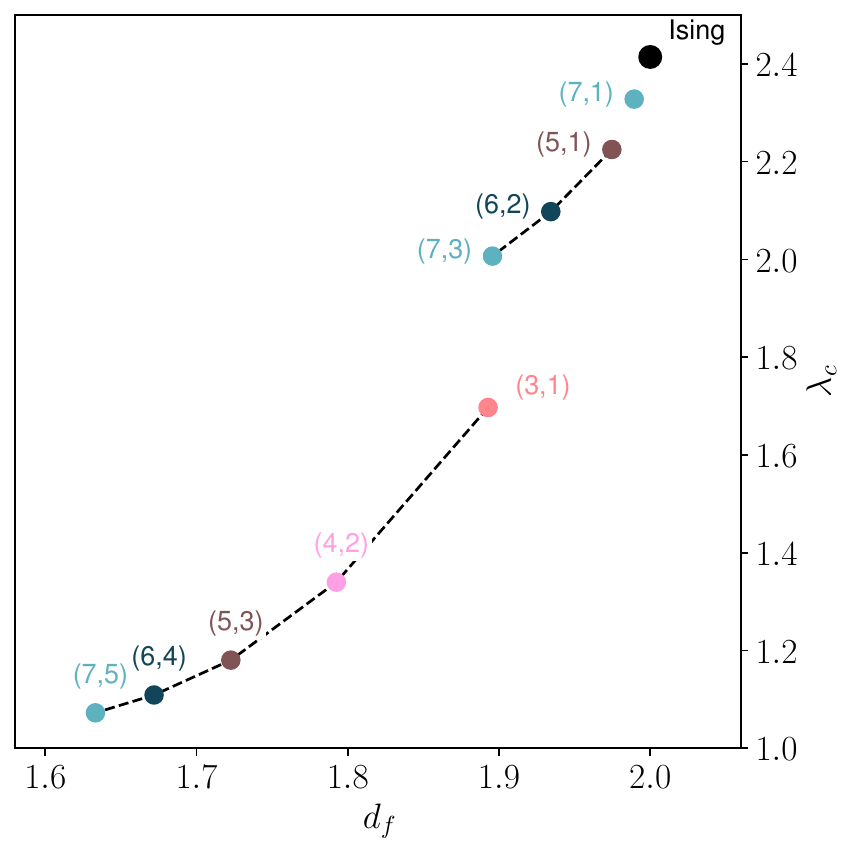}
\caption{
Critical eigenvalue $\lambda_c$ as a function of the fractal dimension $d_f$ for the Sierpiński carpets $SC_k(a,b)$ studied in this work. The data naturally organize into geometric families associated with fixed $(a-b)$. Each family interpolates between the square-lattice Ising limit $d_f=2$ and increasingly filamentary geometries approaching the effectively one-dimensional regime $d_f\to1$. The lowest family, containing the canonical carpet $SC_k(3,1)$, exhibits the slowest convergence and the most demanding computational scaling.
}
\label{pattern}
\end{figure}

Fig.~\ref{pattern} reveals a nontrivial organization of the critical eigenvalues $\lambda_c$ as a function of the fractal dimension $d_f$.
Rather than collapsing onto a single universal curve, the data naturally organize into geometric families associated with fixed values of $(a-b)$. In particular, the regular Sierpiński carpets may be grouped into sequences of the form $SC_k(2n+m,m)$, where $n$ labels the family while $m$ parametrizes the individual members within it. For fixed $n$, these families interpolate between
the square-lattice limit $m=0$, corresponding to $d_f=2$ and increasingly filamentary, gasket-like, geometries approaching an effectively one-dimensional regime in the limit $m\to\infty$, where $d_f\to1$.

The present data therefore suggest that the critical behavior of Sierpiński carpets is not determined uniquely by the fractal dimension alone, but also depends on additional geometric characteristics of the construction, such as connectivity and lacunarity. Interestingly, the lowest family --- containing the canonical carpet $SC_k(3,1)$ --- also exhibits the slowest numerical convergence together with the most demanding computational scaling.
By contrast, the higher families converge substantially faster and display behavior progressively closer to the ordinary
two-dimensional Ising model. This observation indicates that the present study probes precisely the most nontrivial region currently accessible within our computational resources.

% SC42 / SC51 / SC53 / SC62 — single-column float, stacked vertically
\begin{table}[h!]
\centering\small
\caption{Data for the carpet $SC_k(4,2)$.}
\label{tableSC42}
\resizebox{\linewidth}{!}{%
\begin{tabular}{rccrr}
\toprule
$k$ & $\lambda_c$ & $T_c$ & $L$ & \hphantom{5\,159\,780\,3}\# elem. \\
\midrule
2 & 1.6041705 & {\bf 1.3689132} & 16      & 1\,344 \\
3 & 1.5060656 & {\bf 1.2501539} & 64      & 15\,744 \\
4 & 1.4533974 & {\bf 1.1845113} & 256     & 187\,392 \\
5 & 1.4206009 & {\bf 1.1428005} & 1\,024  & 2\,242\,560 \\
6 & 1.3981662 & {\bf 1.1138400} & 4\,096  & 26\,886\,144 \\
7 & 1.3822210 & {\bf 1.0930218} & 16\,385 & 322\,535\,424 \\
8 & 1.3707690 & {\bf 1.0779402} & 65\,536 & 5\,159\,780\,352 \\
\midrule
$\infty$ & 1.3399(5) & {\bf 1.0367(7)} & -- & -- \\
\bottomrule
\end{tabular}}
\\[8pt]
\caption{Data for the carpet $SC_k(5,1)$.}
\label{tableSC51}
\resizebox{\linewidth}{!}{%
\begin{tabular}{rccrr}
\toprule
$k$ & $\lambda_c$ & $T_c$ & $L$ &\hphantom{5\,159\,780\,3} \# elem. \\
\midrule
2 & 2.2369019 & {\bf 2.0789872} & 25      & 6\,564 \\
3 & 2.2276947 & {\bf 2.0690440} & 125     & 157\,236 \\
4 & 2.2255394 & {\bf 2.0667154} & 625     & 3\,772\,164 \\
5 & 2.2250436 & {\bf 2.0661796} & 3\,125  & 90\,524\,436 \\
6 & 2.2249301 & {\bf 2.0660571} & 15\,625 & 2\,172\,548\,964 \\
\midrule
$\infty$ & 2.22489673(23) & {\bf 2.06602096(25)} & -- & -- \\
\bottomrule
\end{tabular}}
\\[8pt]
\caption{Data for the carpet $SC_k(5,3)$.}
\label{tableSC53}
\resizebox{\linewidth}{!}{%
\begin{tabular}{rccrr}
\toprule
$k$ & $\lambda_c$ & $T_c$ & $L$ & \hphantom{5\,159\,780\,3}\# elem. \\
\midrule
2 & 1.4645691 & {\bf 1.1985651} & 25       & 2\,316 \\
3 & 1.3556639 & {\bf 1.0578699} & 125      & 36\,156 \\
4 & 1.2949558 & {\bf 0.97482828} & 625     & 573\,996 \\
5 & 1.2570926 & {\bf 0.92064223} & 3\,125  & 9\,161\,436 \\
6 & 1.2323103 & {\bf 0.88389258} & 15\,625 & 146\,470\,476 \\
7 & 1.2156996 & {\bf 0.85857663} & 78\,125 & 2\,342\,965\,116 \\
\midrule
$\infty$ & 1.18043(3) & {\bf 0.8026(5)} & -- & -- \\
\bottomrule
\end{tabular}}
\\[8pt]
\caption{Data for the carpet $SC_k(6,2)$.}
\label{tableSC62}
\resizebox{\linewidth}{!}{%
\begin{tabular}{rccrr}
\toprule
$k$ & $\lambda_c$ & $T_c$ & $L$ &\hphantom{5\,159\,780\,3} \# elem. \\
\midrule
2 & 2.1151468 & {\bf 1.9468695} & 36     & 11\,376 \\
3 & 2.1017736 & {\bf 1.9322685} & 216    & 363\,168 \\
4 & 2.0987090 & {\bf 1.9289199} & 1\,296 & 11\,616\,192 \\
5 & 2.0980077 & {\bf 1.9281534} & 7\,776 & 371\,687\,040 \\
\midrule
$\infty^*$ & 2.0978051(19) & {\bf 1.9279320(21)} & -- & -- \\
\bottomrule
\end{tabular}}
\end{table}

% SC64 / SC71 / SC73 / SC75 — single-column float, stacked vertically
\begin{table}[t]
\centering\small
\caption{Data for the carpet $SC_k(6,4)$.}
\label{tableSC64}
\resizebox{\linewidth}{!}{%
\begin{tabular}{rccrr}
\toprule
$k$ & $\lambda_c$ & $T_c$ & $L$ &\hphantom{5\,159\,123}  \# elem.\\
\midrule
2 & 1.3747534 & {\bf 1.0832002}  & 36     & 3\,552 \\
3 & 1.2641710 & {\bf 0.93093954} & 216    & 69\,312 \\
4 & 1.2055940 & {\bf 0.84286864} & 1\,296 & 1\,375\,872 \\
5 & 1.1709600 & {\bf 0.78693837} & 7\,776 & 27\,455\,232 \\
6 & 1.1494454 & {\bf 0.7501781}  & 46\,656 & 548\,731\,392 \\
\midrule
$\infty$ & 1.1091(3) & {\bf 0.6753(6)} & -- & -- \\
\bottomrule
\end{tabular}}
\\[8pt]
\caption{Data for the carpet $SC_k(7,1)$.}
\label{tableSC71}
\resizebox{\linewidth}{!}{%
\begin{tabular}{rccrr}
\toprule
$k$ & $\lambda_c$ & $T_c$ & $L$ &\hphantom{5\,159\,123}  \# elem.\\
\midrule
2 & 2.3305762 & {\bf 2.1797575} & 49      & 26\,988 \\
3 & 2.3283611 & {\bf 2.1773824} & 343     & 1\,294\,836 \\
4 & 2.3280153 & {\bf 2.1770116} & 2\,401  & 62\,148\,012 \\
5 & 2.3279620 & {\bf 2.1769544} & 16\,807 & 2\,983\,075\,764 \\
\midrule
$\infty^*$ & 2.3279536(6) & {\bf 2.1769455(6)} & -- & -- \\
\bottomrule
\end{tabular}}
\\[8pt]
\caption{Data for the carpet $SC_k(7,3)$.}
\label{tableSC73}
\resizebox{\linewidth}{!}{%
\begin{tabular}{rccrr}
\toprule
$k$ & $\lambda_c$ & $T_c$ & $L$ & \hphantom{5\,159\,123} \# elem.\\
\midrule
2 & 2.0278300 & {\bf 1.8511724} & 49      & 17\,508 \\
3 & 2.0117047 & {\bf 1.8333998} & 343     & 698\,556 \\
4 & 2.0079611 & {\bf 1.8292690} & 2\,401  & 27\,929\,892 \\
5 & 2.0070846 & {\bf 1.8283016} & 16\,807 & 1\,117\,109\,244 \\
\midrule
$\infty^*$ & 2.00681217(4) & {\bf 1.82800090(4)} & -- & -- \\
\bottomrule
\end{tabular}}
\\[8pt]
\caption{Data for the carpet $SC_k(7,5)$.}
\label{tableSC75}
\resizebox{\linewidth}{!}{%
\begin{tabular}{rccrr}
\toprule
$k$ & $\lambda_c$ & $T_c$ & $L$ & \hphantom{5\,159\,123} \# elem.\\
\midrule
2 & 1.3127905 & {\bf 0.99966188} & 49       & 5\,052 \\
3 & 1.2057708 & {\bf 0.84314562} & 343      & 118\,308 \\
4 & 1.1525460 & {\bf 0.75558932} & 2\,401   & 2\,818\,812 \\
5 & 1.1222604 & {\bf 0.70075026} & 16\,807  & 67\,507\,428 \\
6 & 1.1043108 & {\bf 0.66569732} & 117\,649 & 1\,619\,169\,852 \\
\midrule
$\infty$ & 1.0726(14) & {\bf 0.597(3)} & -- & -- \\
\bottomrule
\end{tabular}}
\end{table}

%───────────────────────────────────────────────────────────────────────────────
\section*{Tilting}
%───────────────────────────────────────────────────────────────────────────────

\begin{figure}%[ht!]
\centering
\includegraphics[width=\columnwidth]{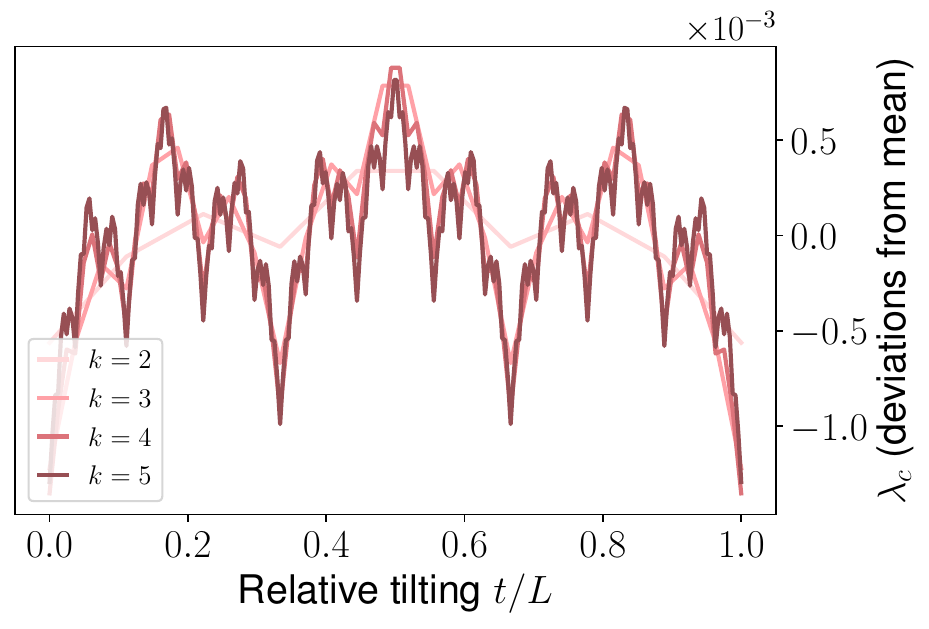}
\caption{Critical eigenvalue $\lambda_c$ dependence on the scaled tilting $t/L$ for the $SC_k(3,1)$ carpet with $k=2,3,4,5$. The deviations from the average over tilting values are actually shown for each generation.}
\label{tilting}
\end{figure}

The approach we have developed relies on making finite approximations (the fractal generations) of the infinite system of interest (the fractal).
This procedure, however, is not unique, and alternative prescriptions can be devised.
A straightforward generalization of the scheme employed here consists in the periodization of the elementary square plaquette with a relative horizontal tilt.
Specifically, one may tile the plane by stacking rows of plaquettes with a relative tilt $t=0,\ldots, L-1$ (in a brick-wall pattern), where $t=0$ corresponds to the standard square-lattice periodization.
The resulting eigenvalues might be as well interpreted as spectral functions probing the appearance of order at different wave vectors (different from $k=0$ mode).
Two further remarks are in order, although they do not apply to Sierpi\'{n}ski carpets.
Vertical and horizontal tilting need not be equivalent in general and, for certain tilting prescriptions the resulting periodic lattice may become disconnected, thereby destroying magnetic order at any finite temperature.
The spectra associated with tilted periodizations are shown in Fig.~\ref{tilting} for $SC_k(3,1)$ at several small generations, as a function of the scaled tilt $t/L$.
As it can be gleaned from this figure, the deviation from the average over tilting appears to converge toward a fractal curve with finite width.
This behavior further illustrates the intrinsic subtlety involved in defining phase transitions---and their associated critical temperatures---on fractal geometries.
Nevertheless, the $t=0$ case, corresponding to the uniform mode in spectral language, emerges as the most natural choice.
It is the periodization that most closely resembles the target infinite fractal and, moreover, it is consistent with previous determinations of the critical temperature.
For these reasons, we adopt the $t=0$ mode as our best estimate for each generation throughout this work.

%───────────────────────────────────────────────────────────────────────────────
\section*{Conclusions and Outlook}
%───────────────────────────────────────────────────────────────────────────────

We have shown that an improvement of the method first presented in Ref.~\cite{Codello:2015bia} is capable
of obtaining accurate critical temperatures for fractal structures with $d_f\leq 2$.
We have focused on the prototypical family of Sierpi'{n}ski carpets for which we provided state-of-the-art estimates of the critical temperature.
While for the original fractal $SC_k(3,1)$ -- for which we reached $k=10$ -- a couple of additional generations are probably still needed to fully settle the extrapolation, for more rapidly converging members of the family -- specifically $SC_k(5,1)$, $SC_k(6,2)$, $SC_k(7,1)$ and $SC_k(7,3)$ -- convergence to several significant digits was already achieved prior to extrapolation.
This demonstrates that, when sufficient computational resources are available, our method is capable of determining the critical temperature with essentially exact precision.
For fractals exhibiting slower convergence than $SC_k(3,1)$, the analysis becomes more challenging due to the steep scaling of the method with the generation $k$.
In these cases, controlled extrapolation remains essential in order to extract reliable estimates of the critical temperature.
An additional outcome of the present analysis is the emergence of a natural geometric organization of regular Sierpi\'{n}ski carpets into families associated with fixed values of $a-b$.
Each family interpolates between the square-lattice Ising limit and increasingly filamentary geometries approaching an effectively one-dimensional regime.
Interestingly, the lowest family -- containing the canonical carpet $SC_k(3,1)$ -- also displays the slowest convergence together with the most demanding computational scaling, indicating that it probes the more interesting regime currently accessible within our present computational resources.
By contrast, the higher families converge substantially faster and exhibit behavior progressively closer to the ordinary two-dimensional Ising model.

Looking forward, a particularly interesting direction for future work is the systematic study of the relationship between the critical temperature $T_c$ and the spectral dimension $d_s$.
Such an analysis could provide valuable insight into the determination of the lower critical dimension of the Ising model and, more generally, into the problem of universality in fractional dimensions below two.

\subsection*{Acknowledgments}
Authors acknowledges compute and financial support from the CSIS grant I+D-2022-22520220100174UD. A.C. also acknowledges financial support from ANII-SNI-2023-1-1013433.
G.G. work is supported by the Deutsche Forschungsgemeinschaft (DFG, German Research Foundation) under Germany's Excellence Strategy EXC2181/1-390900948 (the Heidelberg STRUCTURES Excellence Cluster) and by the European Union under GA No. 101077500--QLRNet.

\contacts{%
  \safeemail{riccardo.ben.ali.zinati}{usi.ch} \\
  \safeemail{gori}{thphys.uni-heidelberg.de} \\
  \safeemail{alessandro.codello}{unive.it}%
}

\spcolophon
\bibliography{bib}

@article{gefen_1984aa,
doi = {10.1088/0305-4470/17/6/024},
url = {https://doi.org/10.1088/0305-4470/17/6/024},
year = {1984},
month = {apr},
publisher = {},
volume = {17},
number = {6},
pages = {1277},
author = {Y Gefen and A Aharony and B B Mandelbrot},
title = {Phase transitions on fractals. III. Infinitely ramified lattices},
journal = {Journal of Physics A: Mathematical and General}
}

@article{gefen:1980aa,
  title = {Critical Phenomena on Fractal Lattices},
  author = {Gefen, Yuval and Mandelbrot, Benoit B. and Aharony, Amnon},
  journal = {Phys. Rev. Lett.},
  volume = {45},
  issue = {11},
  pages = {855--858},
  numpages = {0},
  year = {1980},
  month = {Sep},
  publisher = {American Physical Society},
  doi = {10.1103/PhysRevLett.45.855},
  url = {https://link.aps.org/doi/10.1103/PhysRevLett.45.855}
}

@article{Barlow,
  title={Brownian motion and harmonic analysis on Sierpinski carpets},
  author={Barlow, Martin T and Bass, Richard F},
  journal={Canadian Journal of Mathematics},
  volume={51},
  number={4},
  pages={673--744},
  year={1999},
  publisher={Cambridge University Press},
    doi = {https://doi.org/10.4153/CJM-1999-031-4}
}

@article{vezzani:2003aa,
  title={Spontaneous magnetization of the Ising model on the Sierpinski carpet fractal, a rigorous result},
  author={Vezzani, Alessandro},
  journal={Journal of Physics A: Mathematical and General},
  volume={36},
  number={6},
  pages={1593},
  year={2003},
  publisher={IOP Publishing},
    doi = {10.1088/0305-4470/36/6/305},
    url = {https://doi.org/10.1088/0305-4470/36/6/305}
}

@article{Shinoda_2002,
  title={Existence of phase transition of percolation on Sierpi{\'n}ski carpet lattices},
  author={Shinoda, Masato},
  journal={Journal of applied probability},
  volume={39},
  number={1},
  pages={1--10},
  year={2002},
  publisher={Cambridge University Press},
  doi={https://doi.org/10.1239/jap/1019737983}  
}

@article{Vodvicenko_1965,
  title={A calculation of the partition function for a plane dipole lattice},
  author={Vdovichenko, Natalya V},
  journal={Soviet Physics JETP},
  volume={20},
  number={2},
  pages={477--479},
  year={1965}
}

@book{Feynman_1972,
  title={Statistical mechanics: a set of lectures},
  author={Feynman, Richard P},
  year={2018},
  publisher={CRC press}
}

@article{Codello:2015bia,
doi = {10.1088/1742-5468/2015/11/P11008},
url = {https://doi.org/10.1088/1742-5468/2015/11/P11008},
year = {2015},
month = {nov},
publisher = {IOP Publishing and SISSA},
volume = {2015},
number = {11},
pages = {P11008},
author = {Codello, Alessandro and Drach, Vincent and Hietanen, Ari},
title = {Approximating the Ising model on fractal lattices of dimension less than two},
journal = {Journal of Statistical Mechanics: Theory and Experiment}
}

@article{Perreau,
  title = {Ising model in planar lacunary and fractal lattices: A path counting approach},
  author = {Perreau, Michel},
  journal = {Phys. Rev. B},
  volume = {96},
  issue = {17},
  pages = {174407},
  numpages = {14},
  year = {2017},
  month = {Nov},
  publisher = {American Physical Society},
  doi = {10.1103/PhysRevB.96.174407},
  url = {https://link.aps.org/doi/10.1103/PhysRevB.96.174407}
}

@book{ARPACK,
  title={ARPACK Users' Guide: Solution of Large-Scale Eigenvalue Problems with Implicitly Restarted Arnoldi Methods},
  author={Lehoucq, R. B. and Sorensen, D. C. and Yang, C.},
  year={1998},
  publisher={SIAM},
  address={Philadelphia, PA},
  doi={10.1137/1.9780898719628}
}

@article{Bonnier_1988,
  title = {Real-space renormalization-group study of fractal Ising models},
  author = {Bonnier, B. and Leroyer, Y. and Meyers, C.},
  journal = {Phys. Rev. B},
  volume = {37},
  issue = {10},
  pages = {5205--5210},
  numpages = {0},
  year = {1988},
  month = {Apr},
  publisher = {American Physical Society},
  doi = {10.1103/PhysRevB.37.5205},
  url = {https://link.aps.org/doi/10.1103/PhysRevB.37.5205}
}

@article{Hsiao,
  title = {Critical behavior of the ferromagnetic Ising model on a Sierpi\ifmmode \acute{n}\else \'{n}\fi{}ski carpet: Monte Carlo renormalization group study},
  author = {Hsiao, Pai-Yi and Monceau, Pascal},
  journal = {Phys. Rev. B},
  volume = {67},
  issue = {6},
  pages = {064411},
  numpages = {7},
  year = {2003},
  month = {Feb},
  publisher = {American Physical Society},
  doi = {10.1103/PhysRevB.67.064411},
  url = {https://link.aps.org/doi/10.1103/PhysRevB.67.064411}
}

@article{Genzor2016,
  title = {Phase transition of the Ising model on a fractal lattice},
  author = {Genzor, Jozef and Gendiar, Andrej and Nishino, Tomotoshi},
  journal = {Phys. Rev. E},
  volume = {93},
  issue = {1},
  pages = {012141},
  numpages = {5},
  year = {2016},
  month = {Jan},
  publisher = {American Physical Society},
  doi = {10.1103/PhysRevE.93.012141},
  url = {https://link.aps.org/doi/10.1103/PhysRevE.93.012141}
}

@article{Genzor2019,
  title = {Local and global magnetization on the Sierpi\ifmmode \acute{n}\else \'{n}\fi{}ski carpet},
  author = {Genzor, Jozef and Gendiar, Andrej and Nishino, Tomotoshi},
  journal = {Phys. Rev. E},
  volume = {107},
  issue = {4},
  pages = {044108},
  numpages = {9},
  year = {2023},
  month = {Apr},
  publisher = {American Physical Society},
  doi = {10.1103/PhysRevE.107.044108},
  url = {https://link.aps.org/doi/10.1103/PhysRevE.107.044108}
}

@article{Genzor2023,
  title = {Calculation of critical exponents on fractal lattice Ising model by higher-order tensor renormalization group method},
  author = {Genzor, Jozef},
  journal = {Phys. Rev. E},
  volume = {107},
  issue = {3},
  pages = {034131},
  numpages = {11},
  year = {2023},
  month = {Mar},
  publisher = {American Physical Society},
  doi = {10.1103/PhysRevE.107.034131},
  url = {https://link.aps.org/doi/10.1103/PhysRevE.107.034131}
}

@article{Bonnier:1987aa,
	author = {{Bonnier, B.} and {Leroyer, Y.} and {Meyers, C.}},
	title = {Critical exponents for Ising-like systems on Sierpinski carpets},
	DOI= "10.1051/jphys:01987004804055300",
	url= "https://doi.org/10.1051/jphys:01987004804055300",
	journal = {J. Phys. France},
	year = 1987,
	volume = 48,
	number = 4,
	pages = "553-558",
}

@article{Monceau:1998aa,
  title = {Magnetic critical behavior of the Ising model on fractal structures},
  author = {Monceau, Pascal and Perreau, Michel and H\'ebert, Fr\'ed\'eric},
  journal = {Phys. Rev. B},
  volume = {58},
  issue = {10},
  pages = {6386--6393},
  numpages = {0},
  year = {1998},
  month = {Sep},
  publisher = {American Physical Society},
  doi = {10.1103/PhysRevB.58.6386},
  url = {https://link.aps.org/doi/10.1103/PhysRevB.58.6386}
}

@article{Carmona:1998aa,
  title = {Critical properties of the Ising model on Sierpinski fractals: A finite-size scaling-analysis approach},
  author = {Carmona, Jos\'e M. and Marconi, Umberto Marini Bettolo and Ruiz-Lorenzo, Juan J. and Taranc\'on, Alfonso},
  journal = {Phys. Rev. B},
  volume = {58},
  issue = {21},
  pages = {14387--14396},
  numpages = {0},
  year = {1998},
  month = {Dec},
  publisher = {American Physical Society},
  doi = {10.1103/PhysRevB.58.14387},
  url = {https://link.aps.org/doi/10.1103/PhysRevB.58.14387}
}

@article{Monceau:2001aa,
  title = {Critical behavior of the Ising model on fractal structures in dimensions between one and two: Finite-size scaling effects},
  author = {Monceau, Pascal and Perreau, Michel},
  journal = {Phys. Rev. B},
  volume = {63},
  issue = {18},
  pages = {184420},
  numpages = {10},
  year = {2001},
  month = {Apr},
  publisher = {American Physical Society},
  doi = {10.1103/PhysRevB.63.184420},
  url = {https://link.aps.org/doi/10.1103/PhysRevB.63.184420}
}

@article{Pruessner:2001aa,
  title = {Monte Carlo simulation of an Ising model on a Sierpi\ifmmode \acute{n}\else \'{n}\fi{}ski carpet},
  author = {Pruessner, G. and Loison, D. and Schotte, K. D.},
  journal = {Phys. Rev. B},
  volume = {64},
  issue = {13},
  pages = {134414},
  numpages = {10},
  year = {2001},
  month = {Sep},
  publisher = {American Physical Society},
  doi = {10.1103/PhysRevB.64.134414},
  url = {https://link.aps.org/doi/10.1103/PhysRevB.64.134414}
}

@article{Monceau_2004,
title = {Direct evidence for weak universality on fractal structures},
journal = {Physica A: Statistical Mechanics and its Applications},
volume = {331},
number = {1},
pages = {1-9},
year = {2004},
issn = {0378-4371},
doi = {https://doi.org/10.1016/j.physa.2003.09.045},
url = {https://www.sciencedirect.com/science/article/pii/S037843710300904X},
author = {Pascal Monceau and Pai-Yi Hsiao}
}

@article{Bab:2005aa,
  title = {Critical behavior of an Ising system on the Sierpinski carpet: A short-time dynamics study},
  author = {Bab, M. A. and Fabricius, G. and Albano, E. V.},
  journal = {Phys. Rev. E},
  volume = {71},
  issue = {3},
  pages = {036139},
  numpages = {9},
  year = {2005},
  month = {Mar},
  publisher = {American Physical Society},
  doi = {10.1103/PhysRevE.71.036139},
  url = {https://link.aps.org/doi/10.1103/PhysRevE.71.036139}
}

@article{BAB2009370,
title = {Critical exponents of the Ising model on low-dimensional fractal media},
journal = {Physica A: Statistical Mechanics and its Applications},
volume = {388},
number = {4},
pages = {370-378},
year = {2009},
issn = {0378-4371},
doi = {https://doi.org/10.1016/j.physa.2008.10.029},
url = {https://www.sciencedirect.com/science/article/pii/S0378437108008789},
author = {M.A. Bab and G. Fabricius and E.V. Albano}
}

@article{Jang_2024,
doi = {10.1088/1742-5468/ad0a91},
url = {https://doi.org/10.1088/1742-5468/ad0a91},
year = {2024},
month = {jan},
publisher = {IOP Publishing},
volume = {2024},
number = {1},
pages = {013201},
author = {Jang, Hoseung and Azhari, Mouhcine and Yu, Unjong},
title = {Monte Carlo study for the thermodynamic and dynamic phase transitions in the spin-S Ising model on Sierpiński carpet},
journal = {Journal of Statistical Mechanics: Theory and Experiment}
}

\end{document}